\documentclass[letterpaper, 10 pt, conference]{ieeeconf}

\IEEEoverridecommandlockouts
\usepackage{times}
\usepackage{fancyhdr}
\usepackage{float}
\usepackage{cite}
\usepackage{listings}
\usepackage{booktabs}
\usepackage{subfig, bm}

\usepackage{amsthm}
\usepackage{tikz}
\usepackage{amssymb}
\usepackage{dsfont}
\usepackage{mdwlist}
\usepackage{hyperref}

\hypersetup{colorlinks,linkcolor={blue},citecolor={blue}} 
\usepackage{amsmath,graphicx,xcolor}
\usepackage{subfiles}
\usepackage{import}
\usepackage{dsfont}
\usepackage{xfrac}
\usepackage[font=small,labelfont=bf]{caption}
\usepackage{mathtools}
\usepackage{algorithm}
\usepackage{algorithmic}
\usepackage{adjustbox}
\usepackage{pifont}
\usepackage{sjmacros}
\usepackage{thmstyles}
\usepackage{stfloats}
\usepackage{cuted}
\usepackage{color}
\begin{document}

\title{\LARGE \bf Advantages of Feedback in Distributed Data-Gathering\\ for Accurate and Power-Efficient State-Estimation
}

\author{Hyeongmin Choe$^{1*}$ and SooJean Han$^1$
\thanks{$^1$School of Electrical Engineering, Korea Advanced Institute of Science \& Technology (KAIST), Daejeon, S. Korea.}
\thanks{$^{*}$Corresponding author; Email: \texttt{choe5500@kaist.ac.kr}}
}

\maketitle

\thispagestyle{empty}
\pagestyle{empty}
\allowdisplaybreaks

\begin{abstract}
In distributed target-tracking sensor networks, efficient data gathering methods are necessary to save communication resources and assure information accuracy. 
This paper proposes a \textit{Feedback (FB)} distributed  data-gathering method which
lets the central unit feed information back to the mobile sensors; each sensor then uses it to cancel redundant transmissions and reduce communication congestion. 
We rigorously compare its performance, in terms of mean-squared error (MSE) and cost of power per sensor, against more conventional Non-Feedback (NF) architectures by evaluating conditions of \textit{feasibility} and \textit{advantage} under different architecture specifications (e.g., communication delay rate, power cost rate, maximum backoff time, sampling period, observation noise).
Here, we defined the advantage as the performance gain achieved by FB over NF, while FB is said to be feasible if the advantage region is nonempty. 
Our theoretical analyses show that the feasibility of FB depends more on the 
communication power cost, while the advantage depends on the sensors'
propagation delay per transmission interval; we derive concrete
conditions under which these outcomes hold. 
Using extensive numerical simulations under a variety of settings, we confirm the accuracy of the derived conditions, and show that our theoretical results hold even for more complex scenarios where the simplifying assumptions no longer hold.
\end{abstract}
%

\section{Introduction}
Distributed sensor networks are widely used for target tracking and environmental monitoring where timely and accurate state estimation must be achieved under communication and power constraints (see \cite{castanedo13}  for survey). Conventional architectures typically rely on one-way communication from the sensors to a central unit. But in large-scale networks with overlapping sensing regions and non-negligible communication delays, these architectures often have redundant transmissions among sensors, inefficient power use, and increased estimation error due to large transmission delays~\cite{zhao2002distributed,liu2007survey}.

To address these challenges, \textit{feedback}-enabled network architectures have been studied~\cite{han2022distributed}, in which the central unit feeds information back to the sensors to guide how and when they should transmit new measurements. Such feedback mechanisms allow for dynamic sensor tasking, selective activation, and suppression of redundant data, thereby improving both communication efficiency and estimation quality~\cite{stankovic2005feedback, fatima2023multi}.
Previous literature~\cite{bo2015distributed, han2022distributed, x2014distributed} has empirically demonstrated the benefits of including feedback under
various design constraints, such as limited bandwidth, communication latencies, and partial observability of the environment.
However, determining
explicit conditions under which feedback-based architectures actually outperform traditional sensor networks without feedback remains largely unexplored.

This paper proposes a novel modular feedback (FB) distributed data-gathering architecture that can be used to concretely analyze its performance advantages over more conventional non-feedback (NF) architectures.
Each sensor uses this feedback to cancel redundant data and adjust its transmission schedule based on local rules that aim to reduce its power consumption while delivering its measurements to the central processor as quickly as possible.

Our contributions are as follows. 
First, we derive criteria for \textit{feasibility}, which determines whether there exists a region where FB can have a performance advantage over NF under the same network parameters (communication delay rate, power cost ratio, etc.). 
Second, we rigorously analyze when data-gathering under FB is \textit{advantageous} compared to NF, where no communication occurs from the central unit to the sensors. 
To achieve this, we compare competing performance metrics, highlighting tradeoffs between accuracy (measured by the mean-squared estimation error) and efficiency (measured by the sensors' power cost). 
We support our analyses with extensive simulations across various scenarios, demonstrating that the theoretical benefits of FB persist even when simplifying assumptions are relaxed.

The paper is organized as follows. Section~\ref{problem setup} presents the problem setup. Section~\ref{feedback_arch} introduces the modular architectures used to model both FB and NF.
Section~\ref{theoretical analysis} develops the theoretical results for both feasibility and advantage, while Section~\ref{simulations} validates the theory using simulations. Section~\ref{conclusion} concludes the paper.


\section{Problem Setup}\label{problem setup}
We consider a distributed sensor network which collects positional information about multiple targets, and a central unit that estimates the targets' positions based on received data from the network.
For concreteness of our discussion, we assume $\Scal{\,\subseteq\,}\Rbb^2$ (a planar environment) and that each target moves with a basic period of $T_e{\,\in\,}\Rbb^{+}$, though we emphasize that our results can be easily extended to general aperiodic settings.
For $i{\,=\,}1,\cdots,N$, target $i$'s position, $\xvect_i\in\Scal$, evolves according to a random walk given by
\begin{equation}\label{dynamics}
\xvect_i(t)=\xvect_i(0)+\sum_{k=0}^{\infty}\delta^k(t)\dvect_i^k,\quad t\in\Rbb^+
\end{equation}
where $\delta^k(t)$ is an indicator function that takes value $1$ if $t{\,\in\,}[kT_e, \infty)$, and $\{\dvect_i^k\}_k{\,\in\,}\Rbb^2$ are realizations of i.i.d. random variable stepsizes with each component $d_{i1}^k$ and $d_{i2}^k$ taking value $\Delta d$ with probability $p$, $-\Delta d$ with probability $1-p$.
The full environment state is the stack of all target positions $\xvect{\,=\,}(\xvect_1^{\top},\cdots,\xvect_N^{\top})^{\top}$.

Our network consists of $M$ sensors, where each sensor $j$ has sampling period $T_s{\,\in\,}\Rbb^{+}$ and its own observation region $\Rcal_j{\,\subset\,}\Scal$.
Each sensor $j$ measures the environment state $\xvect$ via $\yvect_j{\,=\,}(\yvect_{j1}^{\top},\cdots,\yvect_{jN}^{\top})^{\top}$, where
\begin{equation}\label{measurement}
    \yvect_{ji}^k=O_{ji}^k(\xvect_i(kT_s)+\vvect_j^k),\quad k\in\Zbb^+.
\end{equation}
is its observation on target $i$.
Here, $O_{ji}^k{\,\in\,}\Rbb^{2\times 2}$ is the identity matrix $I_2$ if $\xvect_i(kT_s){\,\in\,}\Rcal_j$ and $\mathbf{0}$ otherwise, and $\{\vvect_j^k\}_k{\,\sim\,}\Ncal(0,\sigma^2I_2)$ is an uncorrelated noise process independent of the environment state.
We assume the central unit knows the locations and observation ranges of all sensors at the initial time, prior to deployment.

\begin{remark}
    Since the environment evolves in continuous-time according to (\ref{dynamics}), each sensor $j$ also has a continuous-time measurement $\yvect_j(t)$ ($\yvect_{ji}(t)$).
    However, it is piecewise-constant according to its sampling time, and so throughout this paper, we use the ``discretized'' notation $\yvect_j^k$ ($\yvect_{ji}^k$) for each sampling interval $k$ (i.e., $t{\,\in\,}[kT_s, (k+1)T_s)$).
\end{remark}


The sensors and the central unit communicate via \textit{packets}. 
We assume that the communication delay and power cost of transmitting a packet scale linearly with the number of individual vector components it contains, where we refer to the $v_i$ of some vector $\vvect{\,=\,}(v_1,\cdots,v_n)^{\top}$ as its \textit{components}.
The per-component delay and power costs are denoted by $\Delta t^u{\,\in\,}\mathbb{R}^{+}$ and $\Delta p^u{\,\in\,}\mathbb{R}^{+}$ for sensor-to-central (uplink) transmissions, and by $\Delta t^d{\,\in\,}\mathbb{R}^{+}$ and $\Delta p^d{\,\in\,}\mathbb{R}^{+}$ for central-to-sensor (downlink or feedback) transmissions. 


\section{Proposed Feedback Architecture}\label{feedback_arch}

\begin{figure*}[t]
\centerline{\includegraphics[width=0.8\linewidth]{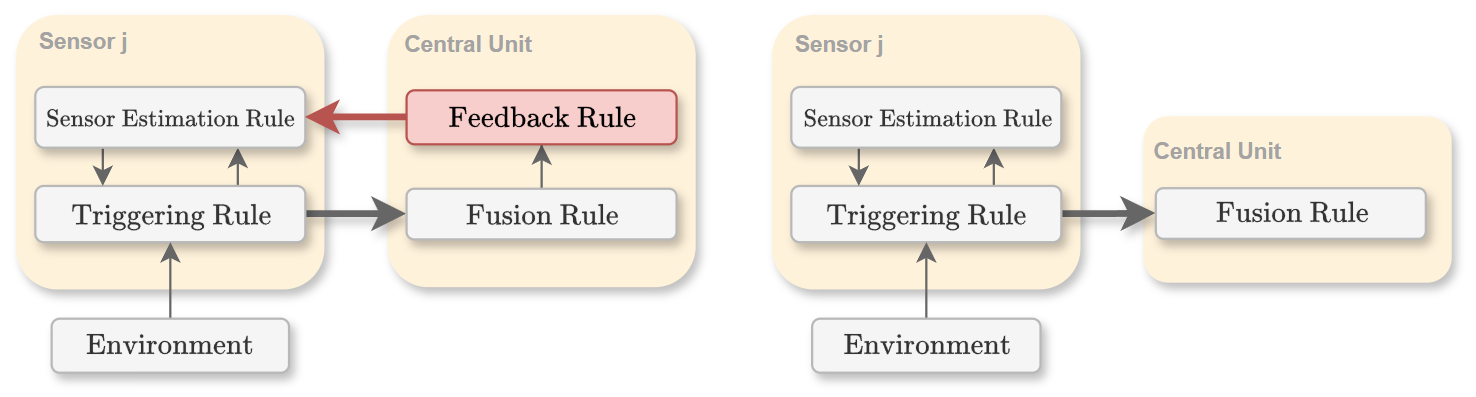}}
\caption{
The modular feedback (FB, left) and non-feedback (NF, right) architectures.  
Each sensor observes the environment and uses a triggering rule (Def.\ref{Triggering Rule def},\ref{transmission rule def}) to decide when to transmit. The transmitted data is fused at the central unit via the fusion rule (Def.\ref{fusion rule def}), and updates the global state estimate.
Under FB, the central unit additionally sends its state estimate as feedback (red, Def.\ref{feedback rule def}) to all sensors. 
Each sensor uses this feedback to update its internal estimate of the central processor's estimate via the estimation rule (Def.\ref{zvect def}).
Note that NF lacks this feedback loop; sensors rely solely on local information to create its internal estimate.
}
\label{algorithm_comparison}
\vspace{-.3cm}
\end{figure*}

The overall objective of the distributed sensor network is to get the central unit to construct $\hat{\xvect}$, an estimate of the environment's state $\xvect$, using only the measurements $\{\yvect_1,\cdots,\yvect_M\}$ transmitted to it by the sensors.
We especially aim to quantify the advantages of adding feedback capabilities to the central unit so that it can communicate information to the sensors, potentially reducing the amount of transmissions they need to send to ensure estimation accuracy.
Towards this goal, we design and compare two architectures: 1) the \textit{feedback (FB) architecture}, which includes feedback, and 2) the \textit{``non-feedback'' (NF) architecture}, which is more commonly used in sensor fusion applications.

As illustrated in Figure~\ref{algorithm_comparison}, the only distinction between FB and NF lies in the presence of the feedback rule, which enables the central unit to transmit information back to the sensors. All other components—the estimation, triggering, and fusion rules—are kept identical across both architectures.
While additional mechanisms such as consensus-based fusion~\cite{olfati05}, and distributed scheduling~\cite{toptal09} can enhance distributed estimation performance, our focus is solely on the impact of feedback. To isolate its effect, we deliberately adopt a minimal system design and use identical rules in both FB and NF, where applicable.

For a concrete comparison, we define each rule as follows.
In order for each sensor $j$ to make informed decisions about what to transmit, it maintains its \textit{own estimate} $\zvect_{ji}{\,\in\,}\Rbb^2$ of the \textit{central processor's estimate} $\hat{\xvect}_i$ of target $i$'s position.
We provide a formal definition of $\zvect_j{\,\triangleq\,}(\zvect_{j1}^{\top},\cdots,\zvect_{jN}^{\top})^{\top}$ later in Definition~\ref{zvect def}, after defining the other rules and their notations.
\begin{definition}(Triggering Rule I)\label{Triggering Rule def}
    If target $i$ is detected by sensor $j$ at time $t$, the \textit{triggering rule} evaluates    
    \begin{equation}\label{triggering rule equation}
        \alpha_{ji}^{(1)}(t)=
        \mathds{1}\{\lVert \zvect_{ji}(t)-\yvect_{ji}^{k*}\rVert>\epsilon\}
    \end{equation}
    where $k^*{\,\triangleq\,}\max\{k \in \mathbb{Z} : kT_s < t\}$ and $\epsilon>0$ is a user-chosen change-detection threshold. 
\end{definition}

If multiple sensors transmit their measurements immediately upon being triggered, issues such as false positives and packet collisions may occur. To mitigate this, each sensor schedules a random backoff time, defined as follows.

\begin{definition}[Backoff Times]
    Each sensor $j$ generates a i.i.d. \textit{random backoff time} $b_j^k\sim\text{Unif}(0,T_b]$ if it decides to transmit its measurement $\yvect_j^k$ during sampling step $k$.
    Here, $T_b{\,<\,}T_s$ is the maximum possible backoff time.
\end{definition}
While there are alternative backoff distributions and various other methods for avoiding collisions in the literature (e.g.,~\cite{bender2016,haas03}), we choose backoff times for concreteness and its simple implementation.
\begin{definition}(Triggering Rule II)\label{transmission rule def}
    If~(\ref{triggering rule equation}) yields $1$ for sensor $j$ time $t$, the triggering rule is checked again at time $t+b_{j}^{k^*}$ in case there has been a new update to $\zvect_{ji}$:
    \begin{equation}\label{transmission rule}
        \alpha_{ji}^{(2)}(t{\,+\,}b_{j}^{k^*}) = \mathds{1}\{\alpha_{ji}^{(1)}(t)=1, \lVert \zvect_{ji}(t{\,+\,}b_{j}^{k^*})-\yvect_{ji}^{k*}\rVert > \epsilon\}
    \end{equation}
    If~(\ref{transmission rule}) yields $1$, then sensor $j$ finally transmits $\yvect_{ji}^{k*}$ to the central unit.
\end{definition}
\begin{definition}[Fusion Rule]\label{fusion rule def}
    For both FB and NF, the central unit applies a \textit{fusion rule} to aggregate all received sensor data and update the estimate $\hat{\xvect}(t)$ over time. 
    For concreteness, both FB and NF use a running arithmetic mean over all past received measurements of each component $x_i$ in the packet; we refer to this as \textit{averaging fusion}.
\end{definition}
\begin{definition}[Feedback Rule]\label{feedback rule def}
    FB architectures employ the following \textit{feedback rule} at each time $t$.
    If the estimate for target $i$ has been updated (i.e., $\hat{\xvect}_i(t)\neq\hat{\xvect}_i(t-)$), and if $\hat{\xvect}_i(t)$ lies within the observation range of at least two sensors, then the central unit feeds $\hat{\xvect}_i(t)$ back to the entire sensor network.
    Here, the notation $t-$ is borrowed from math jargon to mean the time ``just before $t$''.
\end{definition}

\begin{definition}(Sensor Estimation Rule)\label{zvect def}
    Each sensor $j$ implements an \textit{estimation rule} to update $\zvect_{ji}$, its local estimate of the central unit's estimate $\hat{\xvect}_i$, for each target $i$.
    Its definition depends on the architecture type.
    For FB, sensor $j$ estimates the central unit's estimate to be the later between the most recent broadcast received and the last transmission it sent:
        \begin{align*}
            \zvect_{ji}^{\texttt{FB}}(t)= \begin{cases}
            \hat{\xvect}_{i}(t^*) &\text{if broadcast is more updated}\\
            \yvect_{ji}^{k^*} &\text{if last transmission is more updated}
        \end{cases}
    \end{align*}
     where $t^*$ is the latest time before $t$ when triggering rule II was equal to $1$, and $k^*{\,\triangleq\,}\max\{k \in \mathbb{Z} : kT_s < t, \alpha_{ji}^{(2)}(kT_s+b_j^k){\,=\,}1\}$.
     In the NF case, $\zvect_{ji}^{\texttt{NF}}(t)=\yvect_{ji}^{k^*}$.
\end{definition}

The performance of the distributed data gathering architecture is evaluated using the \textit{total power cost} (P) and the \textit{mean squared error} (MSE) over the entire simulation duration $T_\text{sim}$.
\begin{subequations}\label{eq:metrics}
    \begin{equation}\label{eq15}
        \mbox{P}\triangleq\int_{0}^{T_{\text{sim}}}\left(\sum^M_{j=1}(P^u_j(t) + P^d_j(t))\right)dt
    \end{equation}
    \begin{equation}\label{eq16}
        \mbox{MSE} \triangleq \frac{1}{T_{\text{sim}}}\int_{0}^{T_{\text{sim}}}\left(\sum_{i=1}^{N}\lVert \xvect_i(t)-\hat{\xvect}_i(t)\lVert^2\right)dt
    \end{equation}
    where $P_j^u(t)$ and $P_j^d(t)$ are the power consumption of sensor $j$ at time $t$ from transmitting and receiving, respectively. 
    Note that $P_j^u\propto \Delta p^u$ and $P_j^d\propto \Delta p^d$ (see end of Section~\ref{problem setup}).
\end{subequations}


\section{Theoretical Analysis}\label{theoretical analysis}
We now theoretically characterize the tradeoff space between the metrics~\eqn{metrics}: the MSE accuracy of $\hat{\xvect}$ versus the total sensor power cost of transmitting $\{\yvect_1,\cdots,\yvect_M\}$ (and receiving $\hat{\xvect}$, in the FB case).
We emphasize that in this section only, we focus on a variety of simplified settings of the original problem setup from Section~\ref{problem setup} for the purpose of deriving concrete results.
Later, in Section~\ref{simulations}, we numerically study our feedback architecture under the original setup (with all simplifying assumptions removed) and show that the results we derive in theory still hold empirically for more complex scenarios.

\subsection{Preliminary Definitions}
Before beginning our comparisons, we introduce several definitions to facilitate our analyses.
Because feedback transmissions are made when targets are detected within a region of the environment shared by at least two sensors, it helps to partition the collective sensor observation regions by ``collaborative sets''.

\begin{figure}[t]
\centerline{\includegraphics[width=1\linewidth]{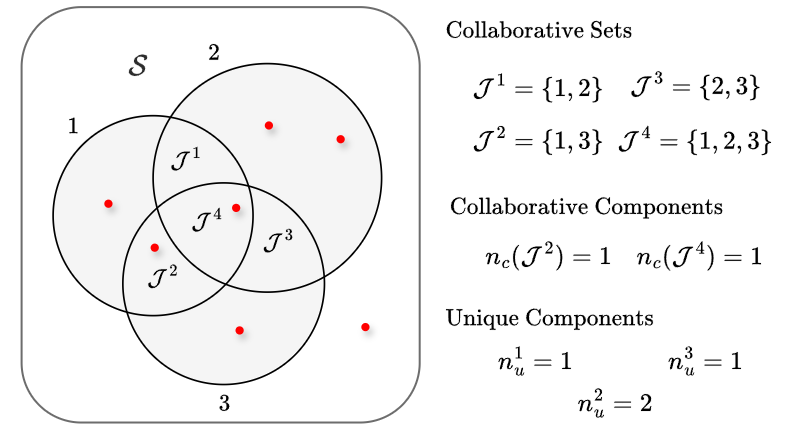}}
\caption{An example illustration of all possible collaborative sets $\{\Jcal_1,\cdots,\Jcal_4\}$ (Def.~\ref{collaborative set def}) and the number of collaborative/unique components (Def.~\ref{def:collab_unique_components}).
This sample environment $\Scal$ (white square) has 3 sensors (gray circles) and 7 targets (red dots).}
\label{collabset}
\vspace{-.3cm}
\end{figure}

\begin{definition}(Collaborative Set)\label{collaborative set def}
    A \textit{collaborative set} $\Jcal=\{j_1,\cdots,j_m\}\subset\{1,\cdots,M\}$, for some $m\leq M$, is created for the sensors $j_1,\cdots,j_m$ iff $\cap_{j\in\Jcal}\Rcal_j$ is a nonempty subset of $\Scal$.
    See Figure~\ref{collabset} for a visualization of all possible collaborative sets for an example of $M{\,=\,}3$ sensors.
\end{definition}

We also consider the distribution of all detected targets in the environment depending on whether they lie in a collaborative set or are uniquely detected by only one sensor.

\begin{definition}[Collaborative and Unique Components]\label{def:collab_unique_components}
    Define, for each collaborative set $\Jcal$, \( n_{\text{c}}(\Jcal) \) to be the number of \textit{collaborative components}, which depends on the number of targets observed in $\cup_{j\in\Jcal}\Rcal_j$.  
    Further define, for each sensor $j$, \( n_{\text{u}}^j \) to be the number of \textit{unique components}, which depends on the number of targets present in $\Rcal_j \backslash \cup_{j'\in\Jcal\setminus\{j\}} \Rcal_{j'}$.
\end{definition}

Each packet incurs power and delay costs proportional to the number of components it contains (see end of Section~\ref{problem setup}). 
Both FB and NF begin with the same number of components due to measuring the same $\{\yvect_1,\cdots,\yvect_M\}$.
However, due to different estimation and triggering rules, FB can let each sensor additionally cancel the transmission of components using the central unit's feedback.
See Figure~\ref{collabset} for an illustration of collaborative and unique components.\\

When a target enters a collaborative set $\Jcal$, each sensor $j \in \Jcal$ responds differently to the trigger rule relative to the others. Under FB, this difference in behavior is further shaped by each sensor’s propagation delay, defined below.

\begin{figure*}[t]
\centerline{\includegraphics[width=0.9\linewidth]{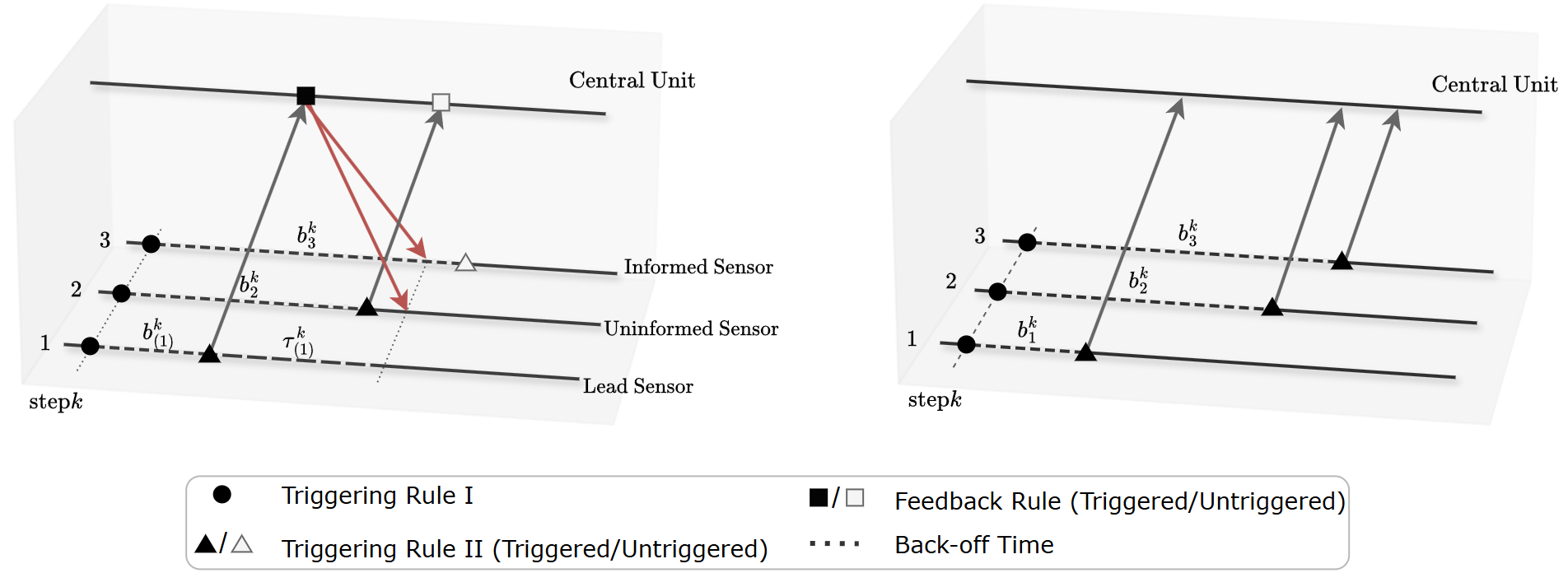}}
\caption{A sample plot of communication events over time for three sensors in the same collaborative set, all triggered simultaneously at step $k$.  
[Left] The lead sensor, sensor 1 ($(1){\,=\,}1$), transmits first (triangle), having the shortest backoff time. 
Under FB, sensor 3 (white triangle) is informed by the feedback and thus does not transmit; sensor 2 (triangle) remains uninformed because it was scheduled during the lead sensor's propagation delay $\tau_{(1)}^k$.
[Right] The equivalent NF case, shown for comparison.}
\label{signal comparison}
\vspace{-.3cm}
\end{figure*}


\begin{definition}[Propagation Delay]\label{propagation delay}
    Under FB only, given a collaborative set $\Jcal$ and sampling step $k$, the \textit{propagation delay} $\tau_j^k$ of each sensor $j\in\Jcal$ is the sum of its transmission delay and feedback delay:
    \begin{equation}
        \tau_{j}^k=n\Delta t^u+m\Delta t^d
    \end{equation} 
    where $n$ is the number of components in the transmission packet of sensor $j$ at time $kT_s+b_j^k$, and $m$ is the number of components in the feedback packet of the central unit at time $kT_s+b_j^k+n\Delta t^u$.
\end{definition}

\begin{definition}(Lead, Informed, and Uniformed Sensors)\label{lead sensorinformed sensors def}
Under FB only, for each collaborative set $\Jcal$, its \textit{lead sensor} at step $k$ is the sensor with the shortest total delay, denoted using the index $(1)$:
\begin{equation}\label{lead criterion}
    (1) \triangleq \arg\min_{j \in \Jcal}( b_j^k+\tau_j^k)
\end{equation}
If more than two sensors have the shortest delay, the lead sensor is the lowest-indexed one.
\textit{Informed sensors} are all sensors in $\Jcal$ that are informed about the environment state via a central processor feedback corresponding to the lead sensor's transmission.
We denote 
$\Ical_k{\,=\,}\{j\in\Jcal:b_j^k{\,>\,}b_{(1)}^k+\tau_{(1)}^k\}$ 
to be the set of all informed sensors in $\Jcal$ at timestep $k$.
Finally, \textit{uninformed sensors} are the remaining sensors in $\Jcal$ which are not lead or informed sensors.
See Figure~\ref{signal comparison} for a visualization of this classification of sensors and delays.
\end{definition}

\subsection{Theoretical Performance Comparisons: FB vs. NF}
We are now ready to compare the FB and NF architectures according to the efficiency and accuracy performance metrics~\eqn{metrics}.
We begin with two preliminary lemmas.

\begin{lemma}\label{powercost difference lemma}
At each step $k$, for each collaborative set $\Jcal$, the power cost NF consumes more power than FB as given by
\begin{equation}\label{pwoercsot difference equation}
\text{P}_k^{\texttt{NF}} - \text{P}_k^{\texttt{FB}} = \sum_{\Jcal}n_c(\Jcal)(\abs{\Ical_k}(\Delta p^u+\Delta p^d)-\abs{\Jcal}\Delta p^d)
\end{equation}
where each $\Ical_k$ is the set of informed sensors of each $\Jcal$ at timestep $k$, and $n_c(\Jcal)$ is the number of components in the collaborative set $\Jcal$ (see Definitions~\ref{lead sensorinformed sensors def},~\ref{def:collab_unique_components}).
\end{lemma}
\begin{proof}
    The power consumption contributed by the unique components
    is identical for both FB and NF architectures.
    For collaborative components $n_c(\Jcal)$,
    the power consumption contribution is  $n_c(\Jcal)\abs{\Jcal}\Delta p^u$ under NF, and $n_c(\Jcal)(\abs{\Jcal}-\abs{\Ical_k})(\Delta p^u+\Delta p^d)$ under FB. Subtracting them yields the desired result~\eqref{pwoercsot difference equation}.
\end{proof}

As we will show in our later theorems, we are interested in deriving conditions for which $\boldsymbol{E}[\text{P}_k^{\texttt{NF}} - \text{P}_k^{\texttt{FB}}]>0$ (and likewise for the MSE).
Note that $\Delta p^u$, $\Delta p^d$, $\abs{\Jcal}$, and related parameters are predetermined via architecture specifications and sensor network layout. 
Thus, a preliminary result that will help us is the expected number of informed sensors given these parameters. 
To obtain interpretable closed-form expressions, we will operate under the following setting.

\begin{assumption}\label{assumption}
    At sampling step $k$, all sensors in collaborative set $\Jcal$ observe the same number of targets.
\end{assumption}
Useful consequences of Assumption~\ref{assumption} are as follows.
Since all sensors share the same targets in $\cup_{j\in\Jcal}\Rcal_j$, the number of unique components $n_\text{u}^j$ is the same for all $j\in\Jcal$.
More importantly, the propagation delays $\tau_j^k$ for all $j\in\Jcal$ are the same.
Therefore, the lead sensor is the sensor whose back-off time is shortest, and~\eqref{lead criterion} is simplified as
\begin{equation}
(1) \triangleq \arg\min_{j \in \Jcal} b_j^k
\end{equation}

\begin{lemma}\label{expected informed lemma}
  Under Assumption~\ref{assumption}, the expected number of informed sensors in collaborative set $\Jcal$ at step $k$, given $\tau_{(1)}^k$, is given by
  \begin{equation}\label{expected informed number}
      \boldsymbol{E}[\abs{\Ical_k}|\tau_{(1)}^k]\propto x^{\abs{\Jcal}}-\abs{\Jcal}x+(\abs{\Jcal}-1)
  \end{equation}
  where $x\triangleq\tau_{(1)}^k/T_b$ is the ratio of the propagation delay of the lead sensor and the back-off interval.
\end{lemma}
\begin{proof}
    As mentioned before, the lead sensor is the sensor whose back-off time is the shortest under Assumption~\ref{assumption}. 
    Using order statistics~\cite{ross96book}, the distribution of the earliest back-off time $b_{(1)}^k$ among a collaborative set is:
    \begin{equation}
        \boldsymbol{P}(b_{(1)}^k=t) = \frac{\abs{\Jcal}}{T_b}\left(1-\frac{t}{T_b}\right)^{\abs{\Jcal}-1}
    \end{equation}
    Given $b^k_{(1)}=t$ and $\tau_{(1)}^k$, the conditional distribution of the number of informed sensors $\abs{\Ical_k}$ is Binomial:
    \begin{equation}
    \begin{aligned}
        \boldsymbol{P}(\abs{\Ical_k}=i| b_{(1)}^k=t,\tau_{(1)}^k) = \binom{\abs{\Jcal}-1}{i}p(t)^i
        \{1-p(t)\}^{\abs{\Jcal}-1-i}
    \end{aligned}
    \end{equation}
    where 
    \begin{equation}
        p(t)\triangleq\frac{T_b-\min(t+\tau^k_{(1)},T_b)}{T_b-t}
    \end{equation}
    The joint distribution can be expressed as follows.
    \begin{equation}\label{eq:powercost_adv_eq0}
        \begin{aligned}
             \boldsymbol{P}(i,t|\tau_{(1)}^k)&=\binom{\abs{\Jcal}-1}{i}p(t)^i
             \{1-p(t)\}^{\abs{\Jcal}-i-1}
             \\&\times
             \frac{\abs{\Jcal}}{T_b}\left(1-\frac{t}{T_b}\right)^{\abs{\Jcal}-1}
        \end{aligned}
    \end{equation}
    where we use shorthand $\boldsymbol{P}(i,t|\tau_{(1)}^k)=\boldsymbol{P}(\abs{\Ical_k}=i, b_{(1)}^k=t|\tau_{(1)}^k)$ for conciseness.
    The expected number of informed sensors can be calculated as follows.
    \begin{align}\label{eq:powercost_adv_eq1}
        \boldsymbol{E}[\abs{\Ical_k}|\tau_{(1)}^k] 
        =\sum^{\abs{\Jcal}-1}_{i=0}i\int_0^{T_b-\tau_{(1)}^k}\boldsymbol{P}(i,t|\tau_{(1)}^k)dt 
    \end{align}
  Substituting~\eqn{powercost_adv_eq0} into~\eqn{powercost_adv_eq1},
    \begin{equation}
    \begin{aligned}
        \boldsymbol{E}[\abs{\Ical_k}|\tau_{(1)}^k] =&\frac{1}{T_b^{\abs{\Jcal}-1}} \sum^{\abs{\Jcal}-1}_{i=0}i\binom{\abs{\Jcal}}{i+1}x^{\abs{\Jcal}-i-1}
        (1-x)^{i+1}  
    \end{aligned}
    \end{equation}
    and by change-of-variables $i+1\mapsto u$,
    \begin{equation}
    \begin{aligned}
        &\boldsymbol{E}[\abs{\Ical_k}|\tau_{(1)}^k] =\\
        &\frac{1}{T_b^{\abs{\Jcal}-1}}\left(\sum^{\abs{\Jcal}}_{u=0}u\binom{\abs{\Jcal}}{ u}x^{\abs{\Jcal}-u}(1-x)^{u}-1+x^{\abs{\Jcal}}\right)
    \end{aligned}
    \end{equation}
    Since the first term is the expectation of $\text{Bin}(\abs{\Jcal},1-x)$, it evaluates to $\abs{\Jcal}x$, and we get the desired result.
\end{proof}

Now we are ready to present our main theorems.
First, we make concrete the notions of \textit{feasibility} and \textit{advantage}. 
The \textit{feasibility} determines whether there exists a configuration where FB architecture can achieve lower power cost than the NF architecture. 
The \textit{advantage} refers to the conditions under which FB outperforms NF under either metric in~\eqn{metrics}.
In this language, for instance, Lemma~\ref{powercost difference lemma} would suggest that the \textit{power cost advantage condition} is equivalent to $\boldsymbol{E}[\text{P}_k^{\texttt{NF}} - \text{P}_k^{\texttt{FB}}]>0$.
Our first theorem rewrites this condition in a way that is more useful in practice, in terms of our system specifications.

\begin{theorem}[Power Cost Advantage]\label{advantage condition theorem}
     Over collaborative set $\Jcal$ and sampling step $k$ satisfying Assumption~\ref{assumption}, given $\tau_{(1)}^k$, FB is \textit{power cost advantageous}  over NF on average if the following conditions are satisfied:
    \begin{subequations}\label{advantage conditions}
        \begin{equation}\label{feasible condition}
            y > \frac{1}{\abs{\Jcal}-1}
        \end{equation}
        \begin{equation}\label{advantage condition}
            (1+y)x^{\abs{\Jcal}} - \abs{\Jcal}(1+y)x + (\abs{\Jcal}-1)y -1 > 0
        \end{equation}
    \end{subequations}
    where \( y \triangleq \Delta p^d / \Delta p^u \)
    and \( x \) was defined in Lemma~\ref{expected informed lemma}, $\tau_{(1)}^k$ is the propagation delay of the lead sensor (see Definition~\ref{propagation delay}).
\end{theorem}
\begin{proof}
   By combining Lemma \ref{powercost difference lemma} and Lemma \ref{expected informed lemma}, we can derive the expected power difference $\boldsymbol{E}[P_k^{\texttt{NF}} - P_k^{\texttt{FB}}|\tau_{(1)}^k]$. Conditioning $\boldsymbol{E}[P_k^{\texttt{NF}} - P_k^{\texttt{FB}}|\tau_{(1)}^k]>0$, we can get the advantage condition~\eqref{advantage condition}. Differentiating $\boldsymbol{E}[P_k^{\texttt{NF}} - P_k^{\texttt{FB}}|\tau_{(1)}^k]$ yields $\abs{\Jcal}(1+y)x^{\abs{\Jcal}-1}-\abs{\Jcal}(1+y)$, which is zero at $x^{\abs{\Jcal}-1}=1$. Since the function is increasing over $(0,1]$, ensuring a nonempty solution set for the advantage condition requires the feasibility condition to hold at $x=0$. This leads to the derived feasibility condition~\eqref{feasible condition}.
\end{proof}

\begin{figure}[t]
\centerline{\includegraphics[width=0.9\linewidth]{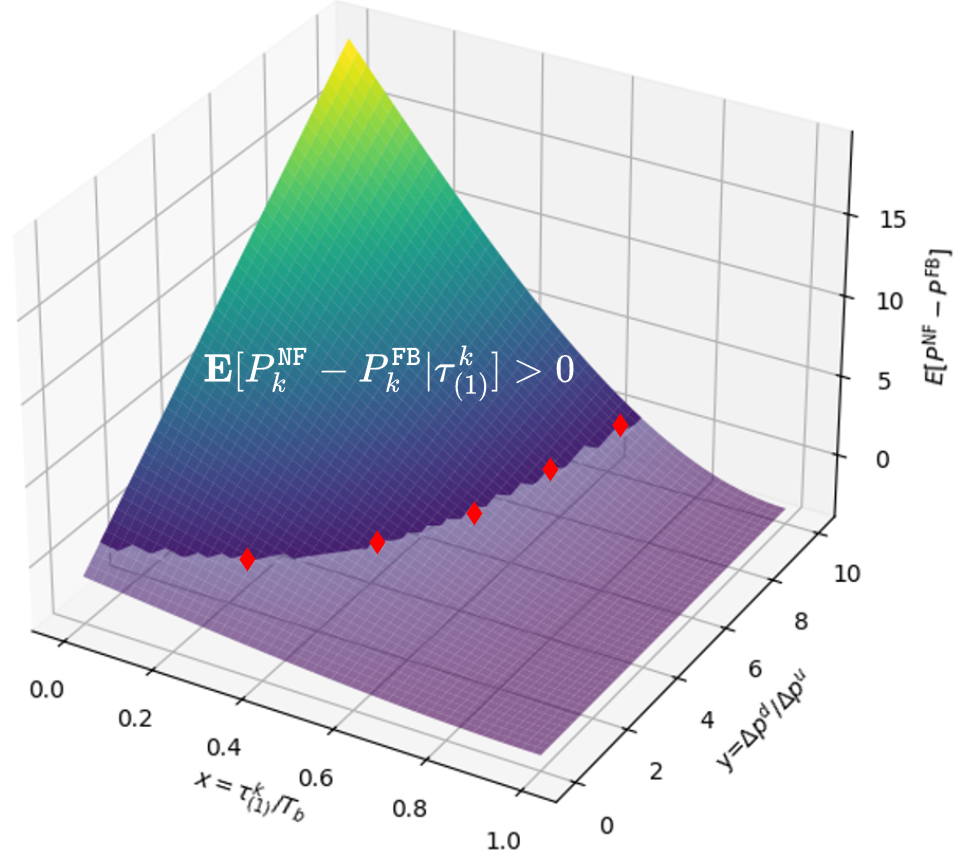}}
\caption{
Expected power cost difference (Lem.~\ref{powercost difference lemma} and~\ref{expected informed lemma}) across \((x, y)\) configurations for an example \( |\mathcal{J}| = 3 \). Here, \( x \) and
\( y \) are defined in Thm.~\ref{advantage condition theorem}.
By the feasibility condition~\eqref{feasible condition}, FB becomes viable when \( y > 0.5 \). 
Moreover, FB is more advantageous according to~\eqref{advantage condition} when the propagation delay is small (\( x \to 0 \))
and $\Delta p^u{\,>\,}\Delta p^d$.
Red dots indicate the boundary where FB and NF yield identical power cost, i.e., the point at which FB starts to become advantageous.
Figure~\ref{various sensor networks} can be viewed as the projection of this type of graph onto the \((x, y)\) plane.
}
\label{advantageous_condition_plot}
\vspace{-.3cm}
\end{figure}

When the feasibility condition~\eqref{feasible condition} is met, FB can outperform NF in terms of power cost. The advantage condition~\eqref{advantage condition} further refines this by identifying the set of states \( x \) where FB is strictly more power-efficient on average. Unlike feasibility, this condition depends on the ratio between the lead sensor’s propagation delay \( \tau_k^{(1)} \) and the backoff interval, both of which may vary with time step \( k \). Figure~\ref{advantageous_condition_plot} explains this visually.

Since both FB and NF aim for the central unit to produce an accurate estimate \( \hat{\xvect} \) of the state \( \xvect \), we also compare them under the MSE metric. The following theorem provides a sufficient \textit{MSE advantage condition} under which FB outperforms (is more advantageous over) NF in terms of the mean-squared error.

\begin{theorem}[MSE Advantage]\label{MSE superior cond}
Let \( n_{\text{u}}^\text{min} \triangleq \min\{n_{\text{u}}^j : j \in \Ical_k\} \) be the minimum number of unique components among the informed sensors, where \( n_{\text{u}}^j \) is defined in Definition~\ref{lead sensorinformed sensors def}. 
Suppose that at every sampling step \( k \), there exists at least one informed sensor \( j \in \Ical_k \) with \( n_{\text{u}}^j > 1 \). Then, FB is \textit{MSE advantageous} over NF if the following condition holds:
\begin{equation}\label{MSE advantage condition}
    \frac{\epsilon}{\sigma} > \sqrt{\frac{2T_s}{\Delta t^u n_{\text{u}}^\text{min}} - 1}
\end{equation}
where \( \epsilon \) is the triggering rules' threshold from Definitions~\ref{Triggering Rule def} and~\ref{transmission rule def}, and \( \sigma \), \( \Delta t^u \), and \( T_s \) are defined in Section~\ref{problem setup}.
\end{theorem}
\begin{proof}
In this proof only, we will enumerate all the collaborative sets and the corresponding sets of informed sensors using the superscript $\ell$, so that $\Jcal\mapsto\Jcal^{\ell}$, $\Ical_k\mapsto\Ical_k^{\ell}$, and $n_{\text{c}}(\Jcal)\mapsto n_c^{\ell}$.
With FB, for each informed sensor \( j \in \Ical_k^{\ell} \) within collaborative set $\Jcal^{\ell}$, the number of scheduled components is reduced by \( n_{\text{c}}^{\ell} \) relative to NF. 
Thus, under FB, each unique component is tracked at least \( \Delta t^u n_{\text{c}}^{\ell} \) faster, and NF accumulates at least
$(\epsilon^2+\sigma^2)n_{\text{u}}^j\Delta t^u n_{\text{c}}^{\ell}$
additional contribution to its MSE compared to FB.
Across the entire sensor network, FB cancels a total of \( \sum_{\ell} \abs{\Ical_k^{\ell}} n_{\text{c}}^{\ell} \) components. The resulting imbalances in MSE are distributed among all informed sensors.  
Thus, the MSE advantage is bounded as:
\begin{equation}
    \text{MSE}_\texttt{adv}\geq (\epsilon^2+\sigma^2) n_{\text{u}}^\text{min} \Delta t^u \sum_{\ell} \abs{\Ical_k^{\ell}} n_{\text{c}}^{\ell}.
\end{equation}
Additionally, since FB has fewer packets in the collaborative set, the corresponding components in the state estimate will have higher variance under FB than NF when both are using the averaging fusion rule of Definition~\ref{fusion rule def} (this is just the law of large numbers).
Thus, we must also account for its MSE \textit{disadvantage}.
Within each collaborative set \( \Jcal^{\ell} \), for each \( n_{\text{c}}^{\ell} \) components, both FB and NF maintain the same MSE until \(\abs{\Jcal^{\ell}} - \abs{\Ical_k^{\ell}}\) identical components—originating from different sensors in $\mathcal{J}^{\ell}$—are collected.  

During this period, the MSE of FB remains either less than or equal to NF due to faster tracking.  
However, beyond this point, FB ceases receiving additional data, resulting in a final error of $\sigma^2/(\abs{\Jcal^{\ell}} - \abs{\Ical_k^{\ell}})$
Conversely, NF continues receiving data, further reducing its error to $\sigma^2/\abs{\Jcal^{\ell}}$.
Thus, NF eventually achieves a lower MSE for the collaborative components.
The imbalance in MSE within a collaborative set persists for at most \(2T_s\), as the backoff time is constrained by the sampling step \( T_b < T_s \).  
Thus, the MSE disadvantage is bounded as:
\begin{equation}
 \text{MSE}_\texttt{dsv} \leq 2T_s \sum_{\ell} \left(\frac{1}{\abs{\Jcal^{\ell}} - \abs{\Ical_k^{\ell}}}-\frac{1}{\abs{\Jcal^{\ell}}}\right)n_{\text{c}}^{\ell} \sigma^2.   
\end{equation}
Considering both advantages and disadvantages, the final condition for FB to be MSE advantageous over NF is $\text{MSE}_\texttt{adv} > \text{MSE}_\texttt{dsv}$, i.e.,
\begin{equation}
 (\epsilon^2+\sigma^2) n_{\text{u}}^\text{min} \Delta t^u \sum_{\ell} \abs{\Ical_k^{\ell}} n_{\text{c}}^{\ell} > 2T_s \sum_{\ell} \frac{\abs{\Ical_k^{\ell}}}{\abs{\Jcal^{\ell}} (\abs{\Jcal^{\ell}} - \abs{\Ical_k^{\ell}})}n_{\text{c}}^{\ell} \sigma^2   
\end{equation}
By assumption, \( \sum_{\ell} \abs{\Ical_k^{\ell}} n_{\text{c}}^{\ell} \neq 0 \) and \( n_{\text{u}}^\text{min} \neq 0 \),
which implies
\[
(\epsilon^2+\sigma^2) > \frac{2T_s}{\Delta t^u n_{\text{u}}^\text{min}} \frac{\sum_{\ell} \frac{\abs{\Ical_k^{\ell}}}{\abs{\Jcal^{\ell}} (\abs{\Jcal^{\ell}} - \abs{\Ical_k^{\ell}})}n_{\text{c}}^{\ell}}{\sum_{\ell} \abs{\Ical_k^{\ell}} n_{\text{c}}^{\ell}} \sigma^2.
\]
Since
$\abs{\Ical_k^{\ell}}/(\abs{\Jcal^{\ell}} (\abs{\Jcal^{\ell}} - \abs{\Ical_k^{\ell}})) \leq \abs{\Ical_k^{\ell}}$ for each $\ell$, we obtain:
\[
(\epsilon^2+\sigma^2) > \frac{2T_s}{\Delta t^u n_{\text{u}}^\text{min}}\sigma^2.
\]
and rearranging the terms yields our desired result.
\end{proof}

From an accuracy standpoint, FB allows each sensor to refrain from transmitting data it believes the central processor has already received,
which accelerates the delivery of new
information; this yields an advantage governed by \( \epsilon \) and \( \Delta t \).
On the other hand, repeatedly fusing redundant data provides a benefit tied to \( \sigma \) and its time of persistence \( T_s \) until the next new sample. 
Overall, FB improves accuracy when the trade-off between these effects favors faster access to new data, i.e., when \( \epsilon / \sigma \) satisfies the condition of Theorem~\ref{MSE superior cond}.

In summary,
FB can offer practical advantages over NF. 
First, by eliminating redundant transmissions, FB reduces power cost under certain conditions (Theorem~\ref{advantage condition theorem}). 
Second, by prioritizing unseen data over redundant ones (Theorem~\ref{MSE superior cond}), FB lowers communication delay, which increases accuracy.


\section{Simulation \& Verification}\label{simulations}
In this section, we use numerical
simulations under more general settings
to assess the practical relevance of our feasibility and advantage conditions,
First, we test whether the qualitative trends predicted by Theorem~\ref{advantage condition theorem} and~\ref{MSE superior cond} persist under realistic conditions where Assumption~\ref{assumption} no longer holds. Next, we validate Theorem~\ref{advantage condition theorem} empirically by 
comparing it with simulation outcomes under settings that do satisfy the assumptions.

\subsection{Feasibility and Advantage over Different Parameters}
\begin{figure*}[t!]
\centerline{\includegraphics[width=\linewidth]{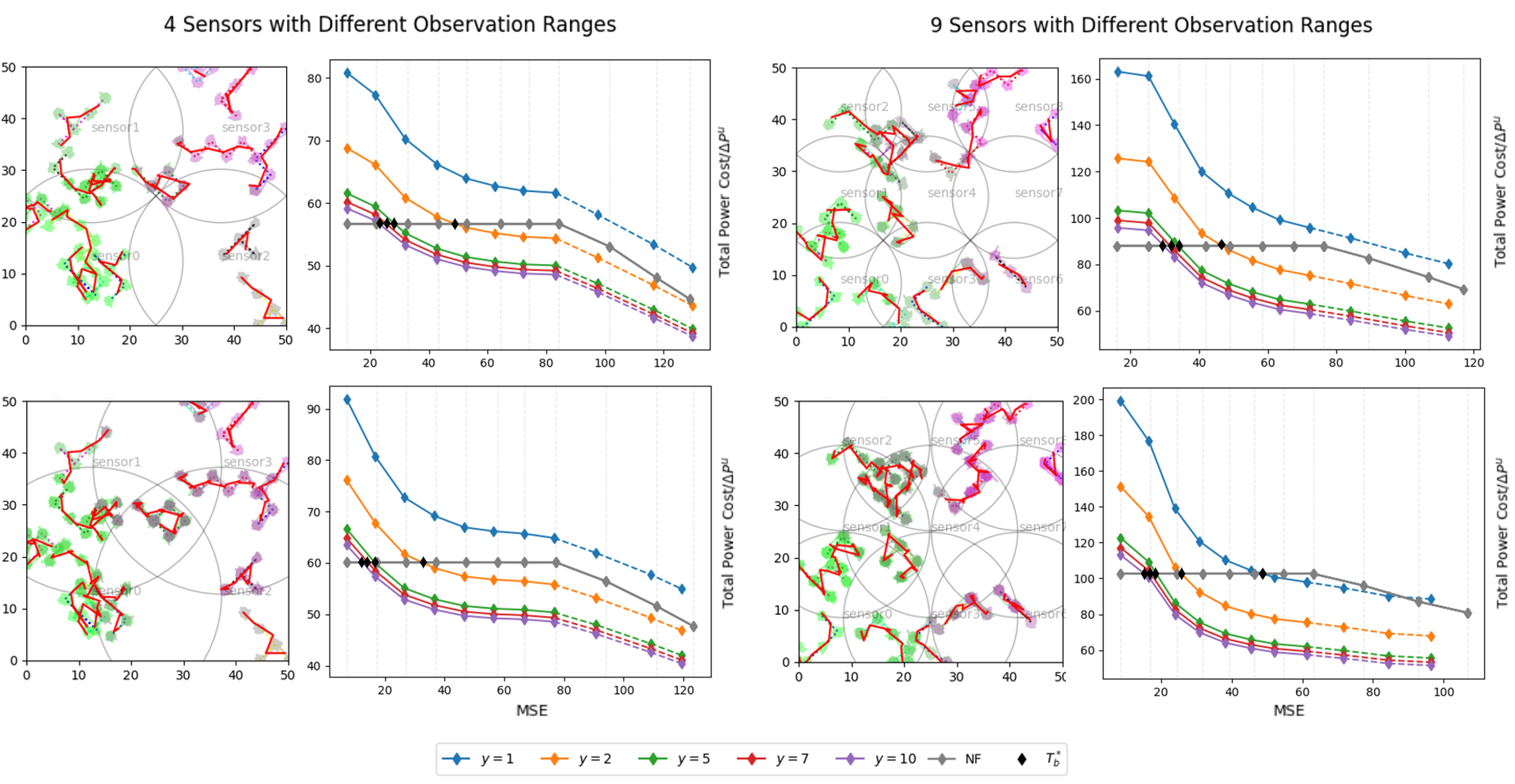}}
\caption{
Simulation results showing MSE versus total power cost under varying \( T_b \) and \( y \), for different sensor networks tracking 15 targets evolving via~\eqref{dynamics}. Each subfigure consists of:
[Left] The sensor environment: gray circles denote observation ranges; colored lines show target trajectories; red lines indicate estimates.  
[Right] MSE vs. total power cost: FB results (colored curves for \( y \in \{1,2,5,7,10\} \)) are compared against NF (gray curve). Colored diamonds represent backoff intervals \( T_b \in \{1,\dots,200\} \), transitioning from solid to dashed when \( T_b > T_s \). Power cost is normalized by \( \Delta p^u \).
While FB consistently achieves lower or equal MSE, it may consume more power than NF at small \( T_b \). Black diamonds mark the crossover point where FB becomes power-efficient. In the top row, \( y=1 \) is often infeasible. However, as observation ranges expand (bottom row) and collaborative regions grow, FB becomes feasible even for \( y=1 \).
}
\label{various sensors}
\end{figure*}

\begin{setup}\label{set:sim1}
The field is defined over the region \( (0,50] \times (0,50] \). A total of 15 targets are initialized at random positions with \( \Delta d{\,=\,}3 \), \( T_e{\,=\,}150 \), \( p{\,=\,}0.5 \). Four and nine sensors with varying observation ranges are deployed as shown in Figure~\ref{various sensors}, with \( T_s {\,=\,} 150 \), \( \sigma {\,=\,} 0.1 \), and \( \epsilon {\,=\,} 2 \).
The per-component communication delays are set to \( \Delta t^u {\,=\,} 2 \) and \( \Delta t^d {\,=\,} 1 \).
To vary \( x \), the backoff interval is swept over \( T_b \in \{1, 20, 40, \ldots, 200\} \). To vary \( y \), the feedback power cost is fixed at \( \Delta p^d {\,=\,} 1 \), while the uplink power cost \( \Delta p^u \in \{1, 2, 3, 4\} \).
Each configuration is evaluated over \( T_{\text{sim}} {\,=\,} 900 \) time steps using 500 Monte Carlo trials.
\end{setup}

\noindent\textbf{Results \& Analysis} \\
We use Setting~\ref{set:sim1} to evaluate the feasibility condition over varying \( y {\,=\,} \Delta p^d / \Delta p^u \), and the advantage condition over \( x {\,=\,} \tau_{(1)}^k / T_b \). Figure~\ref{various sensors} shows the MSE versus total power cost (defined in~\eqn{metrics}) under these variations.

We observe that key behaviors predicted by Theorem~\ref{advantage condition theorem} persist even beyond the simplifying assumptions. For both FB and NF, MSE increases linearly with \( T_b{\,\in\,}[1,140] \) due to backoff delay. However, FB consistently achieves lower MSE by eliminating redundant packet components (see Theorem~\ref{MSE superior cond}).
FB's power cost generally decreases with \( T_b \), except when \( T_b \in [0,20] \), where \( x > 1 \) causes propagation delay to dominate, nullifying the reduction. When \( y = 1 \), FB is never power cost advantageous due to infeasibility. In contrast, for \( y = 2, 3, 4 \), feasibility is ensured, and the point where FB becomes advantageous depends on \( x \).

For large \( T_b \in [140,200] \), both schemes experience reduced power cost, but rapidly growing MSE, as transmission is skipped due to overlap with the next sampling step. Still, FB maintains superior accuracy over NF.

\subsection{Theoretical vs. Empirical Advantage Conditions}\label{subsec:sim2}
\begin{setup}\label{set:sim2}
We adopt the same configuration as in Setting~\ref{set:sim1}, with the following modifications. Three targets move exclusively within a single collaborative region, shared by either \( |\Jcal| = 2 \) or \( |\Jcal| = 3 \) sensors (see the first column of Figure~\ref{various sensor networks}). The target distribution is controlled so that all sensors have a uniform propagation delay of 9, satisfying the specific conditions of Assumption~\ref{assumption}. We set \( T_e = 100 \) and vary the uplink power cost \( \Delta p^u \in \{1, 2, \dots, 10\} \).
\end{setup}
\noindent\textbf{Results \& Analysis} \\
The second column of Figure~\ref{various sensor networks} compares theoretical predictions with empirical results, highlighting regions where FB outperforms NF in power cost as a function of \( T_b \) and \( y \). 
The theoretical model closely matches simulation outcomes. 
Moreover, as \( y \) increases, the advantage region expands consistently, regardless of \( |\Jcal| \).

Increasing the number of collaborating sensors also improves both the feasibility and increases the region where the advantage condition holds. For instance, the advantage region for \( |\Jcal| = 3 \) is not only larger but also reaches higher maximum values than for \( |\Jcal| = 2 \). Notably, when \( y = 1 \), no feasible solution exists for \( |\Jcal| = 2 \), whereas a feasible region exists for \( |\Jcal| = 3 \).

\subsection{More General Sensor Configurations}
\begin{figure*}[t]
\centerline{\includegraphics[width=\linewidth]{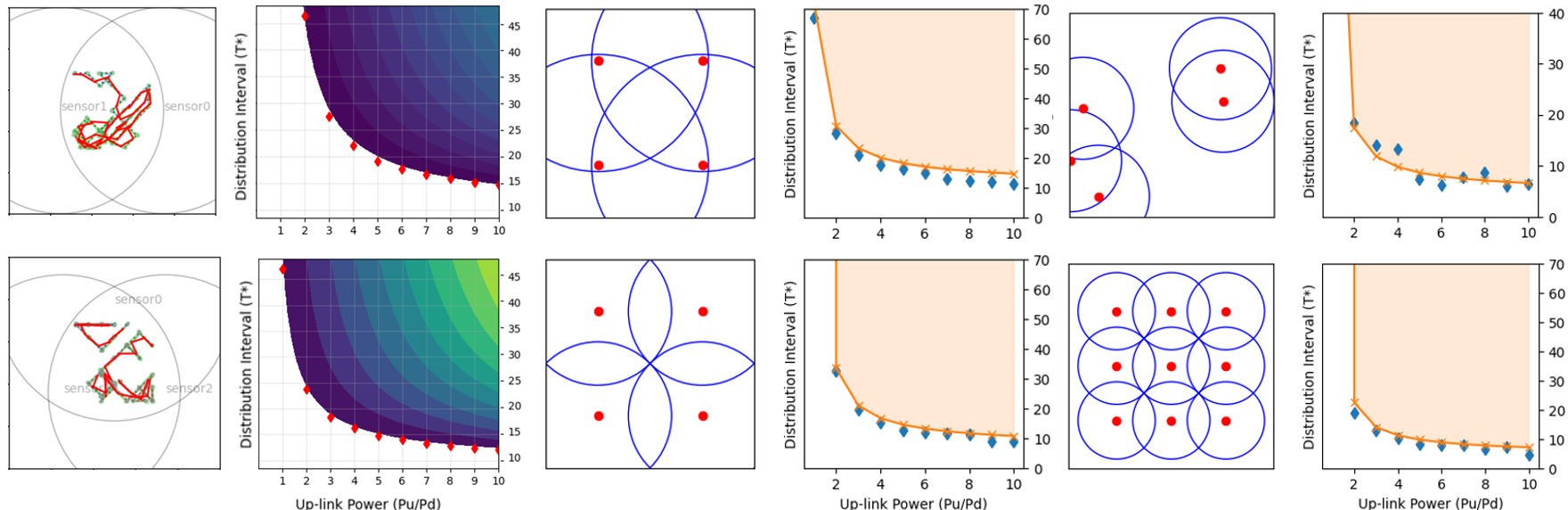}}
\caption{
Theoretical vs. empirical advantage regions across diverse sensor network configurations. In each subfigure, the second column shows advantage regions over \((x, y)\): shaded areas represent predictions from the advantage condition~\eqref{advantage condition}, and dots show simulation results from 500 Monte Carlo trials.
[Columns 1--2] correspond to the controlled setup in Setting~\ref{set:sim2}. The remaining columns reflect approximations under the general setup in Setting~\ref{set:sim3}. In [Columns 3--4], the sensor layouts are identical, but the upper row has larger collaborative sets (\( |\Jcal| \geq 3 \)), making FB feasible even at \( y = 1 \), unlike the lower row where \( |\Jcal| = 2 \) at maximum.
[Columns 5--6, Row 2] also maintain \( |\Jcal| = 2 \), but with many such sensors; thus, the advantage region expands wherever feasibility holds. In [Columns 5--6, Row 1], asymmetric sensor layouts lead to inconsistent packet sizes, reducing the approximation accuracy of our theoretical advantage condition. 
In contrast, symmetric layouts yield closer agreement between theory and simulation due to more uniform averaging.
}
\label{various sensor networks}
\end{figure*}

\begin{setup}\label{set:sim3}
We follow the same configuration as in Setting~\ref{set:sim1}, but with eight targets and diverse sensor layouts (see the third and fourth columns of Figure~\ref{various sensor networks}). The uplink power cost is varied over \( \Delta p^u \in \{1, 2, \dots, 10\} \).
\end{setup}

\noindent\textbf{Results \& Analysis} \\
In this setting, sensor propagation delays vary dynamically. To empirically approximate the advantage region, we applied the advantage condition~\eqref{advantage condition} with the following estimates for each sensor:  
(1) The propagation delay \( \tau_j^k \) was estimated from the ratio of the total observation area to the collaborative area. 
(2) The collaborative set size \( |\Jcal| \) was approximated as the average number of collaborative sets involving the sensor.
Using these estimates, we computed each sensor’s advantage region and averaged them over the entire network. The results, shown in the third to sixth columns of Figure~\ref{various sensor networks}, compare the predicted regions with simulation outcomes, indicating where FB outperforms NF in terms of power cost across \( T_b \) and \( y \).
The trends are consistent with the simpler settings from Section~\ref{subsec:sim2}: larger collaborative areas lead to greater advantage. Also, more symmetric sensor layouts yield tighter approximations, while pronounced asymmetries introduce moderate deviations from the theory.

\section{Conclusion}\label{conclusion}
In this paper, we introduced a modular distributed data-gathering architecture where the central unit is able to feed information back to the sensors.
Each sensor then uses this data to decide how and when to transmit new measurements of the environment to the central unit.
The main benefit of such a feedback (FB) architecture over more conventional non-feedback (NF) architectures is the reduction of the sensors' total power consumption (thereby improving efficiency) and transmission of redundant data, which allows the central unit to attain higher estimation accuracy.
We rigorously derived feasibility and advantage conditions for FB to be able to outperform NF.
Because our conditions related user-chosen parameters with the immutable parameters of the network hardware, they can be used to guide algorithmic design choices and attain a better balance in the accuracy-efficiency trade-off space.
Our conditions were also validated empirically across various parameters, sensor configurations, and environment dynamics, and they were shown to hold even with the simplified theoretical assumptions removed.

This work provides the foundation to a more principled approach to designing scalable and resource-efficient sensor networks for distributed data-gathering and state-estimation problems, particularly in settings with significant communication delays and high sensor density. Future work includes extending the framework to more complex sensor rules, improving the estimation of informed agents, and designing more adaptive backoff mechanisms for faster and more reliable data acquisition.



\bibliographystyle{IEEEtran}
\bibliography{%
refs%
}

\end{document}